# Computing with Spins and Magnets

Behtash Behin-Aein[1], Jian-Ping Wang[2], Roland Wiesendanger[3]

[1]GLOBALFOUNDRIES Inc., Santa Clara, USA, [2]University of Minnesota, Minneapolis, USA
[3]University of Hamburg, Hamburg, Germany

**Key Words:** Magnetic Memory, Gain, Directionality, Non-Volatile Logic, Reconfigurable Logic, Logic-in-Memory, Memory-in-Logic, Spintronics, Magnetic Tunnel Junction, Spin Transfer Torque, Magneto-Resistance, Perpendicular Magnetic Anisotropy, Hysteresis, Spin-Polarized Scanning Tunneling Microscopy, RKKY Interaction.

**Abstract:** The possible use of spin and magnets in place of charge and capacitors to store and process information is well known. Magnetic tunnel junctions are being widely investigated and developed for magnetic random access memories. These are two terminal devices that change their resistance based on switchable magnetization of magnetic materials. They utilize the interaction between electron spin and magnets to read information from the magnets and write onto them. Such advances in memory devices could also translate into a new class of logic devices that offer the advantage of nonvolatile and reconfigurable information processing over transistors. Logic devices having a transistor-like gain and directionality could be used to build integrated circuits without the need for transistor-based amplifiers and clocks at every stage. We review device characteristics and basic logic gates that compute with spins and magnets from the mesoscopic to the atomic scale, as well as materials, integration, and fabrication challenges and methods.

# I. Introduction

Recent experimental advances in spintronics and magnetics have merged the two fields and made possible the use of a single device, namely the magnetic tunnel junction (MTJ), to both read (R) information from magnets and to write (W) information onto magnets.[1] Such devices are being widely researched and developed in industry and academia for spin-based memories. It seems natural to ask whether these advances in memory devices could also translate into a new class of logic devices based on nanoscale magnets or even single spins. What makes logic devices different from memory is the need for such devices to have gain and directionality. Memory bits act as isolated cells for information storage and read out. Logic devices must communicate with each other requiring some bits as inputs and some as outputs. It is essential for inputs to write onto outputs and not the other way around necessitating directionality. Moreover, since logic circuits require multiple cascaded logic bits for computing, information will be lost if not regenerated with gain. Such properties will subsequently enable building of complex circuits for Boolean logic, neural networks, etc. Complementary metal oxide semiconductor (CMOS) devices, which are the current workhorse of transistor-based logic technology, incorporate such characteristics. From this perspective, two categories for spin-based logic exist. In one category, logic is performed in conjunction with transistors and/or clocks[2–8] to provide gain and/or directionality. In the other case, such effects are built into spin devices.[9–16] While covering both categories, our goal here is to present an overview of advancements in computing with spins and magnets from the macroscopic scale to the atomic scale.[16–25]

# II. Magnetic Tunnel Junction

Before we discuss spin-based logic, a brief point about MTJs is in order. MTJs currently act as the most important elemental building block for spin-based memories. As an example, a simplified schematic of an MTJ is shown in Figure 1a. It mainly consists of two magnetic layers (one in dark blue and one in light blue) separated from each other by a thin tunnel barrier made of MgO. The two magnetic layers have different switching thresholds, where one of the layers is utilized as the fixed (reference) layer and the other one as the free (storage) layer. Role of the third magnetic layer (below Ru) is to cancel the stray field of the reference layer on the storage layer. Depending on the parallel (P) or anti-parallel (AP) orientation of the magnetization of the free and fixed layers, the resistance[26,27] (Figure 1a) of the MTJ can change by more than 100%.[27] In addition, the free layer in the MTJ can be electrically switched in a sub-200 ps time scale.[28] The switching energy of the MTJ can be reduced, utilizing different mechanisms such as using a composite structure that preserves the thermal stability while reducing the switching energy.[29] Since magnets hold non-volatile information, the MTJ preserves its state over an electrical power shutdown. All of these properties make the MTJ a promising candidate for logic and memory applications.

### III. Computing with spins and magnets using transistors or clocks

There have been several proposals for utilizing MTJs in computational circuits either as the main core of the computation[2–4] or as a temporary storage element that can hold information, which is called memory-in-logic.[5–7] Figure 2a shows an example of a computational circuit that utilizes an MTJ for its operation and fan out.[6] Here MTJ function as the basic building block for the logic operation and memory unit. Fan out means to transfer the information from one MTJ to another. Once all three input MTJs are in the low resistance (parallel magnetization) state, the current is large enough to switch the magnetization of the output MTJ. Figure 2b demonstrates another computational circuit that works using MTJs.[7] This circuit implements an arithmetic and logic unit. MTJs are connected through a thin and narrow magnetic stripe (or named as nanowire). Magnetic domain walls can be generated and transferred within this magnetic nanowire. The information transfers between MTJs using magnetic domain wall displacement. An actual processor that utilizes MTJs in a logic-in-memory configuration is shown in Figure 2c.[4,5] It is based on 90 nm MTJ/MOS hybrid technology and contains about 0.5 million transistors and 13,400 MTJ cells. In contrast to the conventional electronic, spin information can be transferred over a long distance using dipolar coupling without direct contact of the cells.[30] There have been some proposals for using this dipolar interaction in conjunction with MTJs for computational circuits called magnetic quantum cellular automata (MQCA).[31,32] Figure 2d shows a MTJ-based MQCA that was designed for computation and transfer of information.

### IV. Computing with spins and magnets without transistors or clocks

The approaches discussed thus far involve transistors and/or special clocks[33,36,40] to provide gain and directionality in order to propagate information. We now discuss another approach where such characteristics are inherently built into spin devices.[9,10,13] Figure 3a illustrates a spin switch in which the magnetic free layer, $\hat{m}$, acts as the logic data bit and can orient into the page or out of the page, representing binary 0 or 1. The read process occurs through a double MTJ structure labeled R in the upper dashed box with two fixed magnetic layers $+\hat{M}$ and $-\hat{M}$. Depending on the magnetic orientation of the free layer $\hat{m}$, one MTJ will be in a low resistance (parallel) state, and one will be in a high resistance (anti-parallel) state. This dual MTJ read unit converts positive or negative magnetization (data bits 0 or 1) into a bipolar (that is, positive or negative) output voltage, resulting in a bipolar output current exiting the metal interconnect. Based on the equivalent circuit shown in Figure 3b, the output current can be expressed as:[9,10]

$$I_{\text{out}} = \frac{V \Delta G/G}{R_L + 1/G} \hat{m} \cdot \hat{M}, \qquad (1)$$

where V is the applied voltage which is also the energy supply to the device, $G$ is the sum of the conductances of the two MTJs and $\Delta G$ is their difference. $R_L$ is the load resistance that could be the next spin switch or fan-out in an integrated circuit. This describes the output characteristics (read process) based on the state of the output magnetic data bit $\hat{m}$. The write (W) process occurs in the lower dashed box of Figure 3a which is separated from the read (R) section by an insulator, role of which is to provide electrical input-output isolation.

We know from equation (1) that the state of $\hat{m}$ determines the output. In turn, the switching of this layer is dictated by the free layer $\hat{m}'$ in the (W) section through magnetic coupling. Every time $\hat{m}'$ switches, $\hat{m}$ switches accordingly in order to lower the interaction energy between the two layers that favor the anti-ferromagnetic order to make $\hat{m}$ and $\hat{m}'$ anti-parallel ($\hat{m} || -\hat{m}'$). The switching of $\hat{m}'$ itself is governed by the giant spin Hall effect (GSHE).[34,35] Passing a charge current through a material with high spin–orbit coupling such as Ta drives a spin current into $\hat{m}'$ that can switch its magnetization depending on the polarity of the current.[34,35] The spin Hall effect provides a natural gain.[9,10,36] Without such gain, the output current, instead of the input current, could determine the state of the output, hence eliminating the input-output asymmetry. The equivalent circuit of Figure 3c can be used to calculate gain:[9,10]

$$\text{Gain} \equiv \frac{\Delta I_{\text{out}}}{\Delta I_{\text{in}}} \approx \frac{V\Delta G}{1+R_L G}\frac{\beta}{2I_{s,c}}, \qquad (2)$$

where $I_{sc}$ is the critical spin current needed to switch $\hat{m}'$. Spin current refers to the flow of electron spins that may (spin polarized charge current in MTJs) or may not (pure spin current for GSHE[34,35]) be accompanied by a net non-zero charge current. The symbol β is the ratio of the spin current entering magnet $\hat{m}'$ and the input charge current. For GSHE, as each electronic charge traverses through a material with high spin–orbit coupling, depending on its length, it could emit multiple electron spins to the adjacent magnet. As such, β can be larger than one,[9,10] thus providing gain in the spin switch. However, mechanisms other than GSHE could also prove useful.[12,13,37–39]

The input-output characteristics of the spin switch based on detailed simulations[9,10] is presented in Figure 3b. Hysteresis is present because the magnetic free layer $\hat{m}$ has non-volatile memory. This figure clearly illustrates the presence of gain and input-output asymmetry causing directionality: for a small change in input, there is a large change in output at the transition points, analogous to CMOS devices. One of the simplest circuits where such in-built gain and directionality clearly manifest themselves is that of a ring oscillator. Figure 3d shows an odd number of spin switches connected together. The R unit from each device is connected to the W unit of the next. Since each spin switch is essentially an inverter and there are an odd number of switches, there is no satisfactory steady state that results in an oscillatory output as obtained from detailed simulations[9,10] with no loss of signal strength or compounding errors. The spin switches act autonomously with no need for external transistors or clocks for providing gain and/or directionality in order to propagate information.

Various interconnections of such devices can lead to integrated circuits. Universal Boolean logic gates such as NAND or NOR can be obtained by using spin-based majority logic,[12,15,40] where the output of one device is determined by the majority of the inputs coming from other devices. More interestingly, there are other possibilities beyond standard Boolean logic such as reconfigurable non-volatile circuits[9,10] and neural networks[11] that are enabled with devices that satisfy the essential characteristics discussed previously.

## V. Materials for computing with spins and magnets

In magnetic nanoscale devices, magnetization can be either in the thin-film plane and have in-plane anisotropy (IPA) or normal to the film plane and have perpendicular magnetic anisotropy (PMA). The magnetic thermal stability is usually indicated by the thermal stability factor

$$\Delta \;(= K_{eff}\Omega/k_B T), \tag{3}$$

where $K_{eff}$ is the effective magnetic anisotropy, $\Omega$ is the volume of the magnetic logic/memory cell, $k_B$ is the Boltzmann constant, and $T$ is the temperature. For 10-year data retention, $\Delta$ should exceed 60 at room temperature. Since the magnetic IPA is usually defined by the geometrical aspect ratio of the magnetic nanostructure, magnetic structures with IPA face major thermal stability issues in sub-100 nm dimensions.[41–43] In addition, an IPA-based MTJ requires a large threshold current for spin transfer torque-induced magnetization switching due to the large demagnetization field.[44,45] Here spin transfer torque (STT) refers to the "transferred" torque that is exerted on a nanomagnet by a polarized spin current that passes through the nanomagnet. For a MTJ case, magnetic free layer is the nanomagnet and the polarized spin current is generated by the magnetic fixed layer. In PMA materials, the magnetic anisotropy originates from volume (bulk anisotropy) and interfaces (surface anisotropy). The effective anisotropy energy $K_{eff}$ (Jm$^{-3}$) obeys the relation:[46] $K_{eff} = K_v + \frac{2K_s}{t}$, where $K_v$ (Jm$^{-3}$) is the volume anisotropy and $K_s$ (Jm$^{-2}$) is the interfacial anisotropy. PMA that is induced by interfacial magnetic anisotropy (IPMA) requires a very thin magnetic layer.

There are two classes of IPMA, including bilayer and trilayer structures. Bilayer IPMA has the form of $[X|\{Co,Fe,CoFe\}]_n$, where X is either a heavy transition metal (TM) such as Au, Pd, and Pt, or it is a thin magnetic layer such as Ni. These structures show PMA for certain thicknesses of the magnetic and TM layers.[47–49] This bilayer stack structure needs to be deposited for several periods to provide a strong PMA.[50] This multilayer structure is just like the basic structure of a superlattice of one magnetic layer (e.g. Co) and one non-magnetic layer (e.g. Pd). Furthermore, the PMA can be tuned using the repetition period. There are major drawbacks for this class of bilayer IPMA. The presence of heavy metals lifts their damping constant up to 0.5.[51,52] In addition, the crystalline structure of these materials usually does not match the MgO crystalline structure, and they have a low spin polarization all together, resulting in a low TMR

ratio.[53] There have been some proposed remedies such as by insertion of a thin magnetic layer between the IPMA and the tunnel barrier.[54]

The second class of IPMA is a trilayer structure in the form of X|[Co,Fe,CoFeB]|OX, where X is a heavy element such as Au, Pd, Pt, and Ta, while OX could be any oxide material, such as MgO, $GdO_x$, and $Al_2O_3$.[27,55,56] Among the trilayer structures, Ta|CoFeB|MgO is the most studied structure where a high TMR ratio above 120%[44,57,58] and a decent damping of about 0.015 can be achieved. The $K_{eff}$ of the trilayer structure is $1–2\times10^6$ erg/cc, which fails for the thermal stability of sub-20 nm MTJ.[44] The bulk perpendicular magnetic anisotropy (BPMA) originates from the strong lattice anisotropy field. BPMA materials have a strong $K_{eff}$ that can be as high as $7\times10^7$ erg/cc in FePt films.[59] These alloys incorporate a heavy element, and thus they have a large damping constant on the order of 0.5.[60] BPMA materials usually have a low spin polarization on the order of 50% or less.[61] One solution to the problem of a large damping constant in some promising BPMA materials is to have a composite free layer that consists of a low-damping and high spin polarization magnetic layer exchanged coupled to a BPMA layer.[29]

### VI.    Computing with spins and magnets at the atomic scale

The benefits of scaling electronic circuits are widely known. Higher density reduces cost and improves computational throughput. As such scalability is always an imperative topic for any potential technology. Computing with spins and magnets is not an exception either. An ultimate goal of spintronic research is realizing such concepts at the atomic scale. In fact, it is now possible to characterize materials and their magnetic interactions at these scales. Spin-polarized scanning tunneling microscopy (SP-STM)[17] is a unique enabling tool for transforming spin logic concepts into reality thanks to the unique combination of spin sensitivity,[18] atomic-scale spatial resolution,[19,20] and atom manipulation capability.[21]

As an initial step, a bottom-up approach for fabricating artificial nanomagnets with full control over each constituent atom and their mutual spin-dependent couplings has been developed. The distance dependency of the indirect magnetic exchange (Ruderman-Kittel-Kasuya-Yosida or RKKY) interaction in pairs of magnetic Fe adatoms on Cu(111) was deduced by measuring magnetization curves[22] of each atom in the pairs.[23] The observed oscillatory type of magnetic coupling on the level of individual atoms allowed the tailoring of nanomagnets by controlling the type of pairwise magnetic interactions (either ferro- or antiferromagnetic). A plethora of different nanomagnets ranging from even and odd numbered chains to building blocks of lattices with different symmetries have been realized using tip-induced atom manipulation and simultaneous spin-sensitive imaging (Figure 4). The magnetic ground states of these nanomagnets have then been studied in real space as a function of an external magnetic field by means of single-atom magnetometry using SP-STM.[24] The building blocks of the spin-frustrated lattices show stepwise lifting of multiple degenerate ground states. To reproduce small trends in the magnetization curves of the chains, the correct next-nearest neighbor interactions from an *ab initio* calculation of the full nanomagnet, which slightly differs from the corresponding pairwise interactions, had to be taken into account.

Based on the unique combination of bottom-up atomic fabrication and spin-resolved STM imaging, a prototype system for logical operations has been realized[16] that uses atomic spins of adatoms adsorbed onto a non-magnetic metallic surface and their indirect magnetic exchange interaction in order to transmit and process information (Figure 5). The adatoms have two different states, 0 or 1, depending on the orientation of their magnetization (down or up, respectively). They are constructed to form antiferromagnetically RKKY-coupled chains (as in Figure 4c) that transmit information on the state of small ferromagnetic islands ("input islands") to the gate region. The gate region, which comprises two "input atoms" from each chain and an "output atom," forms the core where the logic operation is performed. The states of the inputs and the resultant state of the output atom are read out by the magnetic tip of the SP-STM (i.e., in the local tunneling magnetoresistance device geometry).[18] Although the STM is used to construct and characterize the device, it is not required to perform the given logic operation. The states of the inputs can be switched independently by external magnetic field pulses. Based on an all-spin concept, this model device is principally non-volatile and functions without the flow of electrons, promising an inherently large energy efficiency.[16] Compared to nanoelectronic devices based on charge transport, contact resistance problems do not occur. Moreover, all-spin atomic-scale devices, such as the one shown in Figure 5 ultimately promise high-speed operation since the time scale for a Fe-adatom spin to flip is only on the order of 200 fs.[25]

## VII. Summary

Computing with spins and magnets offers several advantages if implemented. The information would be non-volatile preventing the logic circuitry from losing information if powered off. This can reduce power dissipation due to leakage for non-active circuitry. Magnetic devices have operating voltages well below 1V[14,62-64] enabling interconnects to operate at lower voltages. This could significantly reduce power consumption in the wiring on the wafer (back end of line) which is a substantial portion of overall power consumption. Computing with spins and magnets can employ majority logic.[9,15] This enables more compact functional gates[65] resulting in increased density. Some logic bits in majority logic can be toggled to reconfigure computational gates making them electrically programmable. Moreover, there have also been proposals for novel computing schemes[11,66] using spins and magnets that could augment current Boolean logic based computing model.

Our objective was to present an overview of how spins and magnets can be used for computing, a necessary step towards implementation of logic circuits that can exploit the advantages mentioned previously. Various state-of-the-art proposals and experiments for spin-based logic applications along with material and fabrication challenges and methods were discussed. The choices of read and write mechanisms presented here are not unique and might change in the future. However, we believe the general topics and concepts we discussed can guide the way for capable spintronic-based logic devices that could enhance computing beyond what is feasible with current technology.

**Acknowledgements**

B. Behin-Aein is indebted to S. Datta for his invaluable advice and consultation. We would also like to thank M. Jamali, B. Chilian, A.A. Khajetoorians, F. Meier, J. Wiebe, A. Klemm, and L. Zhou for their contributions. Financial support from the National Science Foundation (NSF) Nanoelectronics Beyond 2020 (NEB) and the ERC Advanced Grant FURORE and by the Deutsche Forschungsgemeinschaft via the SFB 668 is gratefully acknowledged.

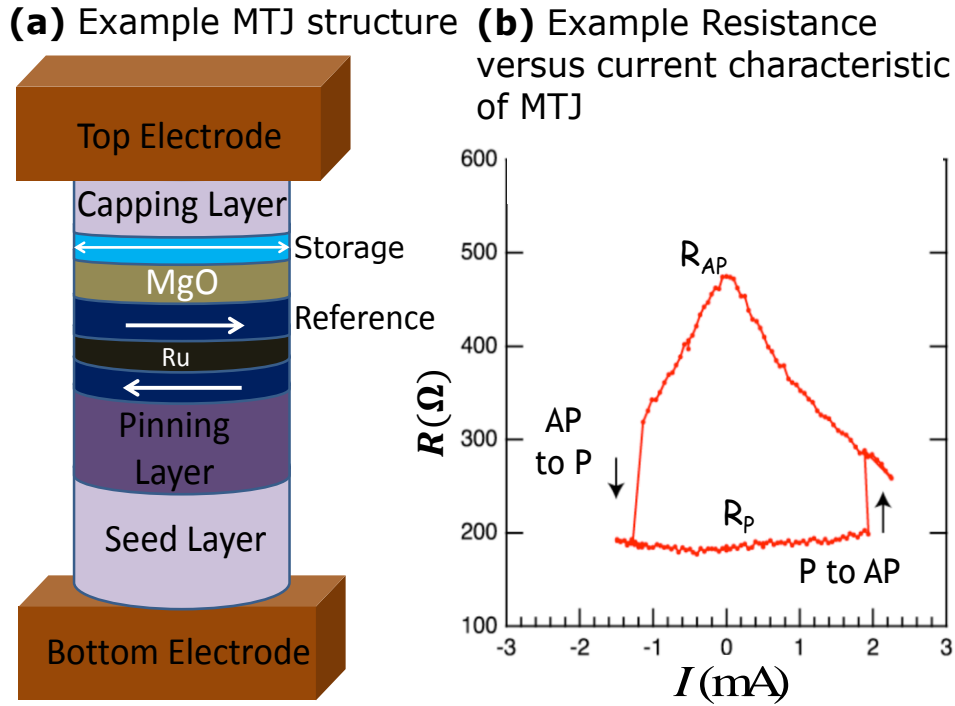

**Figure 1**. (a) A simplified schematic of a magnetic tunnel junction (MTJ). It is mainly consisted of a storage layer and reference layer separated by an MgO layer. Role of the third magnetic layer below Ru is to cancel out the magnetic stray field of the reference layer on the storage layer. (b) Two distinct resistance levels based on the relative orientation of the reference layer and the storage layer are shown when current ($I$) is passed through the structure. Reprinted with permission from Reference 26. © 2008 Macmillan Publishers. Note: AP, antiparallel; P, parallel; $R_{AP}$, resistance in the AP state; $R_P$, resistance in the P state.

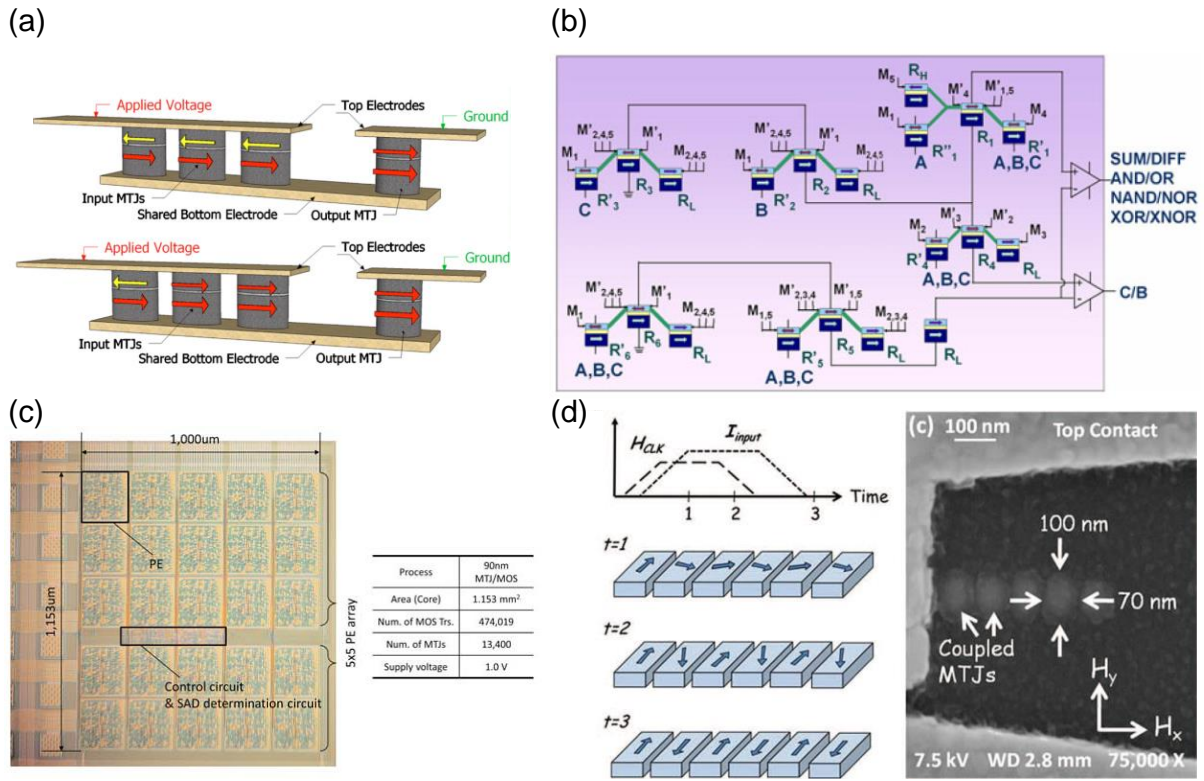

**Figure 2.** (a) A computation circuit fully based on magnetic tunnel junctions (MTJs). 2 The bottom red arrows present the fi xed layer of MTJs, while the top arrows show the free layer of MTJs. Reproduced with permission from Reference 2. © 2010 AIP Publishing LLC. (b) An arithmetic logic unit that utilizes 20 MTJ cells for operation. 7 The main MTJ cells in each logic unit are: $R_1$, $R_2$, $R_3$, $R_4$, $R_5$, and $R_6$. The status of each main MTJ cell is controlled by the input MTJ cells in each logic unit through a nanomagnetic channel (NMC). These input MTJ cells are $R'_1$, $R'_2$, $R'_3$, $R'_4$, $R'_5$, and $R'_6$, which are controlled by one or more current inputs: $A$, $B$, and $C$. $R$ L and $R$ H are reference cells, which have a fi xed low and high resistance, respectively. Each NMC is controlled by one or more control signals: $M_1$– $M'_1$, $M_2$– $M'_2$, $M_3$– $M'_3$, $M_4$– $M'_4$, and $M_5$– $M'_5$. With permission from Reference 7. © 2012 IEEE. (c) A micrograph of a nonvolatile logic-in-memory array processor in 90 nm MTJ/MOS technology[4,5]. The inset table lists elements of the processor, including 474,019 MOS transistors and 13,400 MTJs. The sum of absolute differences (SAD) and processing element (PE) correspond to the SAD and PE circuits. Reprinted with permission from Reference 5. © 2013 IEEE. (d) MTJ-based magnetic quantum cellular automata that utilize the magnetic dipolar interaction for transferring information and for computation. 32 The left part represents the timing of the clock ( $H$ clk ) and the input signal ( $I$ input ) as well as the way the spin signal propagates through the magnetic dipole coupling, while the right one shows a scanning electron micrograph (SEM) of the actual fabricated device. The SEM is taken at a 7.5 kV acceleration voltage for a working distance (WD) of 2.8 mm.

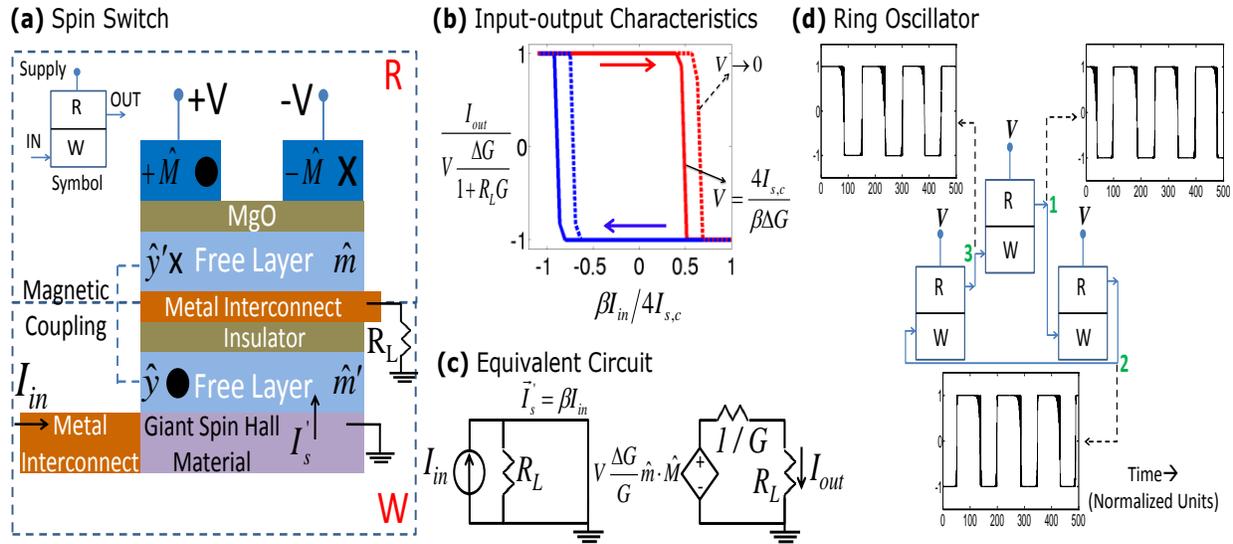

**Figure 3**. (a) A suggested structure for the spin switch composed of read (R) and write (W) units along with a symbol depicting input, output, and energy supply. The y and y' directions are out of the page and into the page respectively. (b) Input-output characteristics of the spin switch derived from (c) the equivalent circuit of spin switch. Directionality and gain are evident because small changes in input result in large changes in output at the edges of the hysteresis. The input signal is amplified with gain as defined in equation (2). These are necessary attributes to make large scale circuits. (d) A simple circuit. An odd number of spin switch devices can be connected to form a ring oscillator similar to complementary metal oxide semiconductor ring oscillators with no loss of signal strength or compounding errors because these devices have gain and directionality. The information is transferred from switch 1 to 2 to 3 and then back to 1 and each switch oscillates. The spin switches act autonomously without the need for external transistors or clocks to propagate information.[9,10] This enables complex and large-scale circuits[9–11] for computation. $V$ is the applied voltage which is the energy supply. $\widehat{M}$ is the magnetization of the two fixed layers that point into and out of the plane. $\hat{m}$ is the magnetization of the free layer acting as the storage layer. $\widehat{m'}$ is the magnetization of the free layer whose orientation controls the orientation of the storage layer. $\hat{y'}$ and $\hat{y}$ are opposite directions into and out of the page and refer to the anti-ferromagnetic orientation of the two free layers. $I_{in}$ is the input current. $I'_S$ is the spin current entering the bottom free layer. $I_{S,C}$ is the critical spin current needed for switching. $I_{out}$ is the outgoing current through $R_L$ which is the load resistance. (This load can be another spin switch or circuitry.) $\beta$ is the ratio (in magnitude) between $I'_S$ and $I_{in}$. $G$ is the sum of the conductances of the two MTJs and $\Delta G$ is the difference.

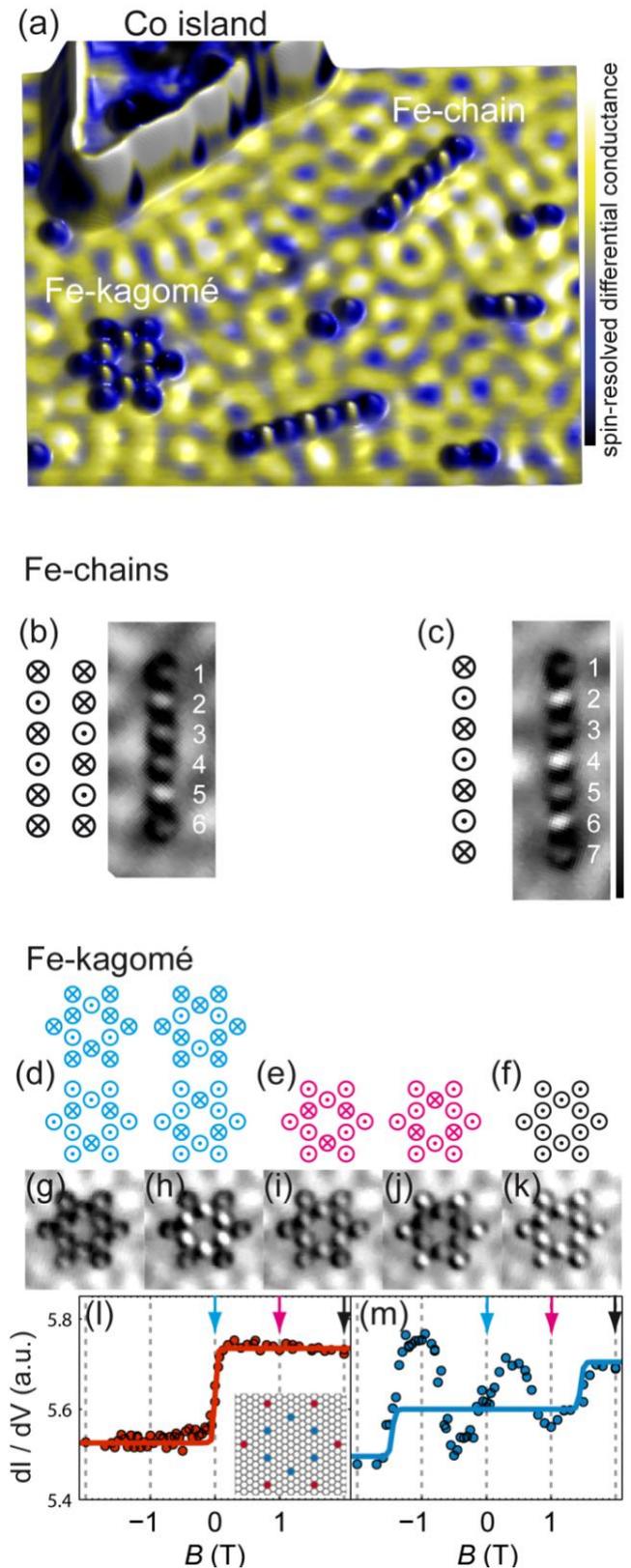

**Figure 4**. (a) Spin-polarized scanning tunneling microscopy (SP-STM) image (25 nm x 20 nm) of fabricated nanomagnets consisting of indirect exchange -coupled Fe atoms on Cu(111). (b–c) Magnetization states from an Ising model (left, partly degenerate) and spin-polarized scanning tunneling spectroscopy () images (right) of chains of antiferromagnetically coupled Fe atoms with a length of six (b) and seven (c) atoms. The distance between individual Fe atoms is 1 nm. (d–f): Degenerate magnetization states from an Ising model for an array of 12 antiferromagnetically coupled atoms at magnetic fields (B), as indicated by the arrows in (l,m). (g–k): Five SP-STM images of a kagomé lattice unit of Fe atoms recorded at B-fields of -2 T, -1 T, 0 T, +1 T, +2 T, respectively, as indicated by the dashed lines in (l,m). The distance between individual Fe atoms is 1 nm. (l,m): Magnetization curves measured on the kagomé atoms (corresponding to the red circles for (l) and the blue circles for (m) in the inset of (l)). Thick colored lines in the magnetization plots show magnetization curves as calculated from an Ising model.[24] Note: $I$: spin-resolved tunneling current ; $V$: applied sample bias voltage .

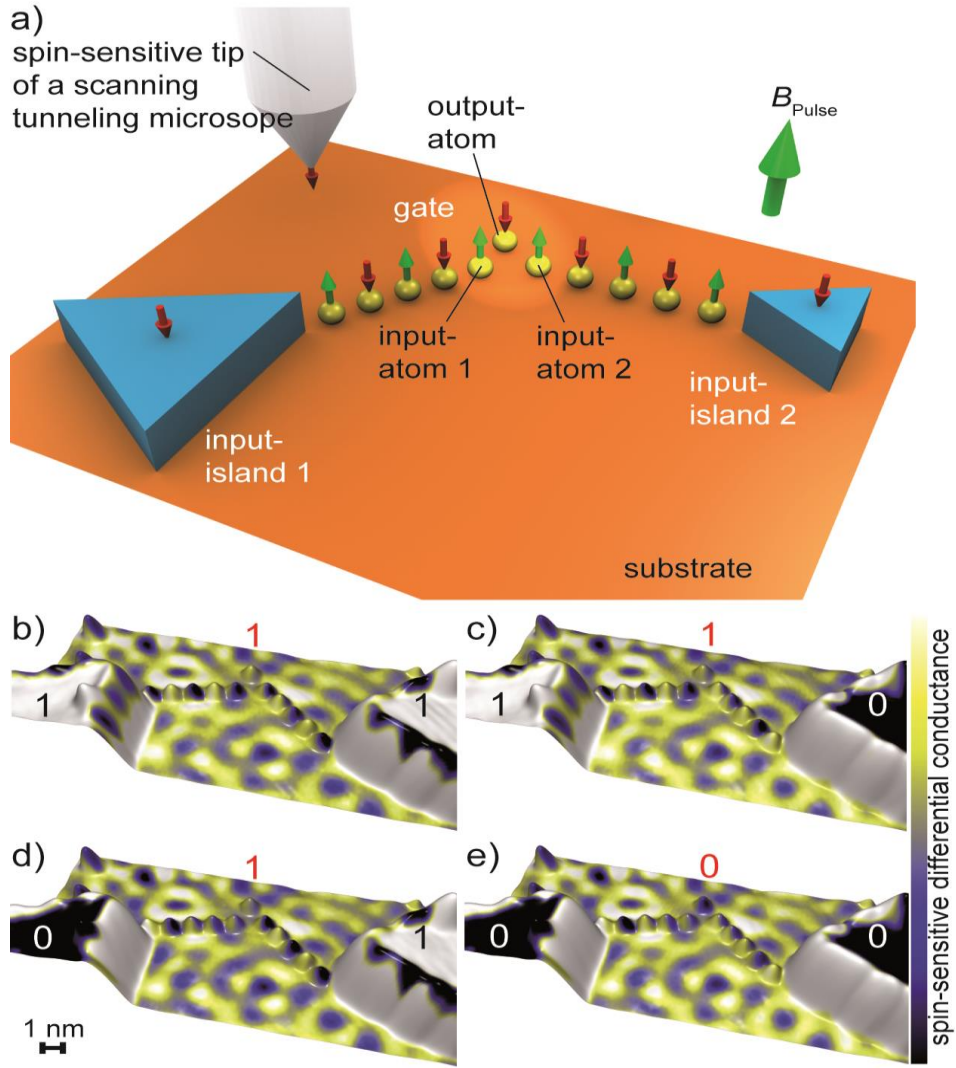

**Figure 5**. (a) Device concept for all-spin atomic-scale logic: Two chains of antiferromagnetically coupled magnetic atoms (yellow spheres) on a nonmagnetic metallic substrate are exchange-coupled to two "input islands" (1, 2) of different size, consisting of patches of ferromagnetic layers. The "input atom" (1, 2) of each spin lead and the final "output atom" form a magnetically frustrated triplet with an antiferromagnetic coupling, which constitutes the logic gate. The field pulse $B_{pulse}$ is used to switch the inputs. The magnetic tip of a scanning tunneling microscope () is used to construct and characterize the device. (b–e) Side-view of 3D topographs colored with simultaneously measured spin-polarized scanning tunneling spectroscopy images of the constructed OR-gate for all four possible input permutations. By applying out-of-plane magnetic field pulses of different strength and direction, each input island can be controllably switched, and the two spin leads transmit the information to their end atoms. The spin state of the output atom in the gate triplet responds accordingly, thereby reflecting the logic operation OR.[16]